\documentclass[useAMS]{mn2e}

\usepackage{graphicx}

\title[Flattened velocity dispersion profiles in GCs] {Flattened velocity dispersion profiles in Globular Clusters: 
Newtonian tides or modified gravity?} 

\author[X. Hernandez, M. A. Jim\'enez, C. Allen] {X. Hernandez, M. A. Jim\'enez, C. Allen\\ 
Instituto de Astronom\'{\i}a, Universidad Nacional Aut\'{o}noma de M\'{e}xico,
  Apartado Postal 70--264 C.P. 04510 M\'exico D.F. M\'exico. \\
}

\date{Released 21/06/2012}

\begin{document}

\label{firstpage}

\maketitle

\begin{abstract}
Over the past couple of years, a number of observational studies have confirmed the flattening of the
radial velocity dispersion profiles for stars in various nearby globular clusters. As the projected
radial coordinate is increased, a radius appears beyond which, the measured velocity dispersion ceases to 
drop and settles at a fixed value, $\sigma_{\infty}$. Under Newtonian gravity, this is explained by invoking
tidal heating from the overall Milky Way potential on the outer, more loosely bound stars, of the globular
clusters in question. From the point of view of modified gravity theories, such an outer flattening is 
expected on crossing the critical acceleration threshold $a_{0}$, beyond which, a transition to MONDian
dynamics is expected, were equilibrium velocities cease to be a function of distance. In this paper
we attempt to sort out between the above competing explanations, by looking at their plausibility 
in terms of an strictly empirical approach. 
We determine Newtonian tidal radii using masses accurately calculated through stellar population modelling, 
and hence independent of any dynamical assumptions, distances, size and orbital determinations for a sample of 16 
globular clusters. We show that their Newtonian tidal radii at perigalacticon are generally larger 
that the radii at which the flattening in the velocity dispersion profiles occurs, by large factors of 4, on average. 
While this point makes the Newtonian tidal explanation suspect, it is found that the radii at which 
the flattening is observed { on average} correlate with the radii where the $a_{0}$ threshold is crossed, and that 
$\sigma_{\infty}$ values scale with the fourth root of the total masses, all features predicted under modified 
gravity theories. 

\end{abstract}


\section{Introduction} \label{intro}

The central values of the stellar velocity dispersion, projected on the plane of the sky, for many Galactic globular clusters 
(GC) have been well known for decades, and are known to accurately correspond to the expectations of self-consistent dynamical models under
Newtonian gravity, e.g. King models (e.g. Binney \& Tremaine 1987, Harris 1996). Recently, a number of studies 
(e.g. Scarpa et al. 2007a, 2007b, 2010 \& 2011 and Lane et al. 2009, 2010a, 2010b \& 2011, henceforth the Scarpa et al. and Lane et al. groups respectively) 
have performed measurements of the projected velocity dispersion along the line of sight for stars in a number of Galactic GCs, but as a function of 
radius, and reaching in many cases out to radial distances larger than the half-light radii of the clusters by factors of a few.

The surprising result of the above studies has been that radially, although velocity dispersion profiles first drop along Newtonian 
expectations, after a certain radius, settle to a constant value, which varies from cluster to cluster. This behaviour is what
is expected under MOND (Milgrom 1983), where equilibrium velocities tend to a constant value when below a critical acceleration, $a_{0}$. In fact,
such a result is fairly generic to modified theories of gravity designed to explain galactic rotation curves in the absence
of any dark matter, e.g. Milgrom (1994), Bekenstein (2004), Zhao \& Famaey (2010), Bernal et al. (2011), Mendoza et al. (2011), Capozziello \& De Laurentis (2011).
As already noted by Scarpa et al. (2011), it is suggestive of a modified gravity scenario
that the point where the velocity dispersion profiles flatten, approximately corresponds to the point where average stellar accelerations
drop below $a_{0}$. Several recent studies have shown dynamical models for self-gravitating populations of stars under MOND or
other modified gravity variants (e.g. Moffat \& Toth 2008, Haghi et al. 2009, Sollima \& Nipoti 2010, Haghi et al. 2011, Hernandez \& Jim\'enez 2012) 
which accurately reproduce not only the observed velocity dispersion profiles, but also the observed surface brightness profiles of observed GCs.

A recent study reaching the same conclusions, but at a significantly distinct scale, can be found in our work Hernandez, Jim\'enez 
\& Allen (2012), where we show that the relative velocity of wide binaries in the solar neighbourhood is in conflict with predictions
from full galactic dynamical simulations of the systems observed, and actually shows also velocities which cease to drop
with distance, precisely on crossing the same $a_{0}$ threshold.
Along the same lines, Lee \& Komatsu (2010) show that the infall velocity of the two components of the Bullet cluster, as required to
account for the hydrodynamical shock observed in the gas, is inconsistent with expectations of full cosmological simulations under
standard $\Lambda CDM$ assumptions. This has recently been confirmed at greater detail by Thompson \& Nagamine (2012), and can in fact be
seen as a failure not only of the $\Lambda CDM$ model, but of standard gravity, as the required collisional velocity is actually
larger than the escape velocity of the combined system. We note also the recent reviews by Kroupa et al. (2010), Famaey \& McGaugh (2012) and Kroupa (2012) and 
references therein, detailing a number of observations in tension with standard $\Lambda CDM$ assumptions. 

From the point of view of assuming Newtonian gravity to be exactly valid at all low velocity regimes, it has also been
shown that both velocity dispersion and surface brightness profiles for Galactic GCs can be self-consistently
modelled. Under this hypothesis, it is dynamical heating due to the overall Milky Way potential that is responsible for the
flattening of the velocity dispersion profiles e.g. Drukier et al. (2008), K\"{u}pper et al. (2010), Lane et al. (2010). 
The constant velocity dispersion observed at large radii merely shows the contribution of unbound stars in the process of evaporating 
into the Milky Way halo. In attempting to sort between these two contrasting scenarios, here we take a fully empirical approach. 
We critically examine the plausibility of both gravitational scenarios by looking through the data for other correlations which each suggest. 

For the Newtonian case, we examine the best available 
inferences for the tidal radius of each cluster at closest galacto-centric passage, and compare it to the observed point
where the velocity dispersion flattens. Here we find the former to generally exceed the latter by factors of 4 on average,
making the Newtonian interpretation suspect. Also, we take all the clusters which the Lane et al. group have claimed 
show no indication of a modified gravity phenomenology, based on the fact that their velocity dispersion profiles can be modelled
using Plummer profiles, and show that the fits with the generic asymptotically flat profiles we use are actually slightly better,
in all cases. 

From the point of view of MONDian modified gravity theories, we reexamine in greater 
detail the correlation between the crossing of the $a_{0}$ threshold and the point where the velocity dispersion flattens,
already suggested by Scarpa et al. (2011). We shall use the term MONDian to refer to any modified theory of gravity which reproduces the 
basic phenomenology of MOND in the low velocity limit for accelerations below $a_{0}$, of flat equilibrium velocities and a Tully-Fisher relation,
regardless of the details of the fundamental theory which might underlie this phenomenology.

In consistency with the expectations of such theories, we find that { mostly}, systems almost
fully within the $a<a_{0}$ threshold show almost fully flat velocity dispersion profiles, while those which only reach this threshold at their
outskirts, present a significant Newtonian region, with a large fall in their velocity dispersion profiles. The above 
correlations are actually what would be expected generically under modified gravity schemes. We confirm our
previous results { with a much smaller sample of} Hernandez \& Jim\'enez (2012), showing that the asymptotic values of the velocity 
dispersion profiles are consistent with scaling with the fourth root of the total masses, a Tully-Fisher relation for GCs. Our results support
the interpretation of the observed phenomenology as evidence for a change in regime for gravity on crossing the $a_{0}$ threshold.

In section (2) we present the detailed velocity dispersion fitting procedure, and show the best fit profiles, including a comparison
with the Plummer models used by the Lane et al. group, which are slightly worse than the asymptotically flat profiles we use.
A description of the tidal radii derivations and the calculation of the confidence intervals for all the globular cluster
parameters used is also given. Section (3) shows a comparison of the tidal radii against the radii at which the velocity
dispersion becomes flat, as a test of the plausibility of Galactic tides under a Newtonian scenario as responsible for
the observed outer flattening of the velocity dispersion profiles. In section (4) we present a number of scalings
between the structural parameters of the observed globular clusters, showing these systems to be consistent with
MONDian gravity expectations, in terms of a change towards a modified regime on crossing the $a_{0}$ threshold.
Our conclusions are summarised in section (5).

\section{Empirical velocity dispersion modelling}

We begin by modelling the observed projected radial velocity dispersion profiles, $\sigma_{obs}(R)$, in the globular clusters in our sample,
listed in table 1. As seen from the Scarpa et al. and Lane et al. data, the observed velocity dispersion profiles show a central core 
region where the velocity dispersion drops only slightly, followed by a ``Keplerian'' zone where the drop is more pronounced. 
These first two regions are in accordance with standard Newtonian King profiles, but they are then followed by a third outermost 
region where the velocity dispersion profiles cease to fall along Keplerian expectations, and settle to fixed values out to the last 
measured point. As some of us showed in Hernandez \& Jim\'enez (2012), an accurate empirical modelling for these velocity dispersion
profiles can be achieved through the function:

\begin{figure*}
\hskip -10pt \includegraphics[width=18cm,height=19cm]{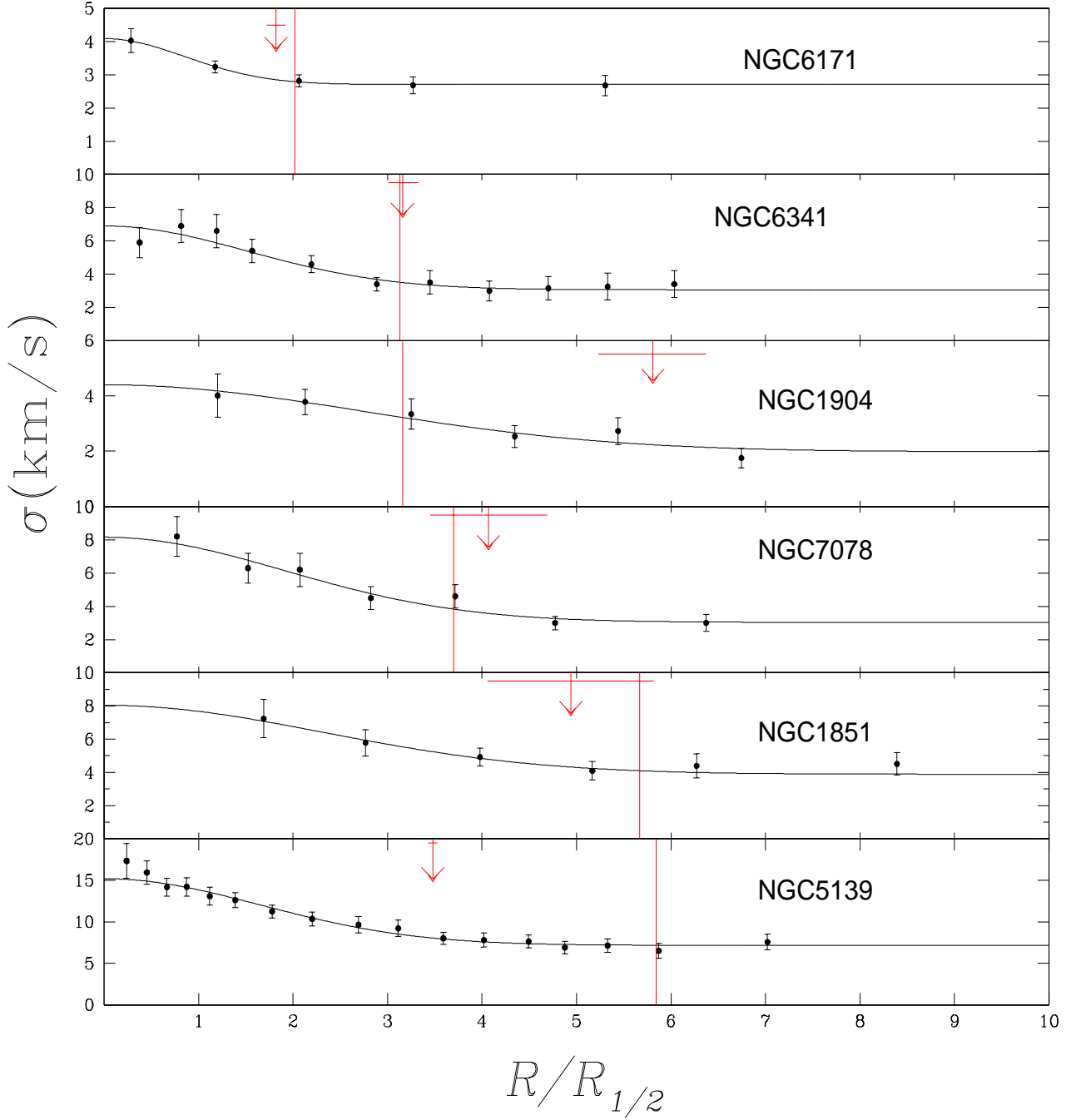}
\caption{The figure shows the observed projected velocity dispersion profiles for six GCs in the Scarpa et al. sample,
points with error bars, as a function of the radial coordinate, normalised to the half-light radius of each. The solid curves 
give the maximum likelihood fits to the asymptotically flat $\sigma(R)$ model of eq.(1), seen to be accurate descriptions of the data. 
The vertical lines indicate the $a=a_{0}$ threshold, and the arrows the point where the profiles flatten, {\it a priori} 
independent features, in most cases seen to occur at approximately the same { region}; see text for details.}
\end{figure*}

\begin{figure*}
\hskip -10pt \includegraphics[width=18cm,height=19cm]{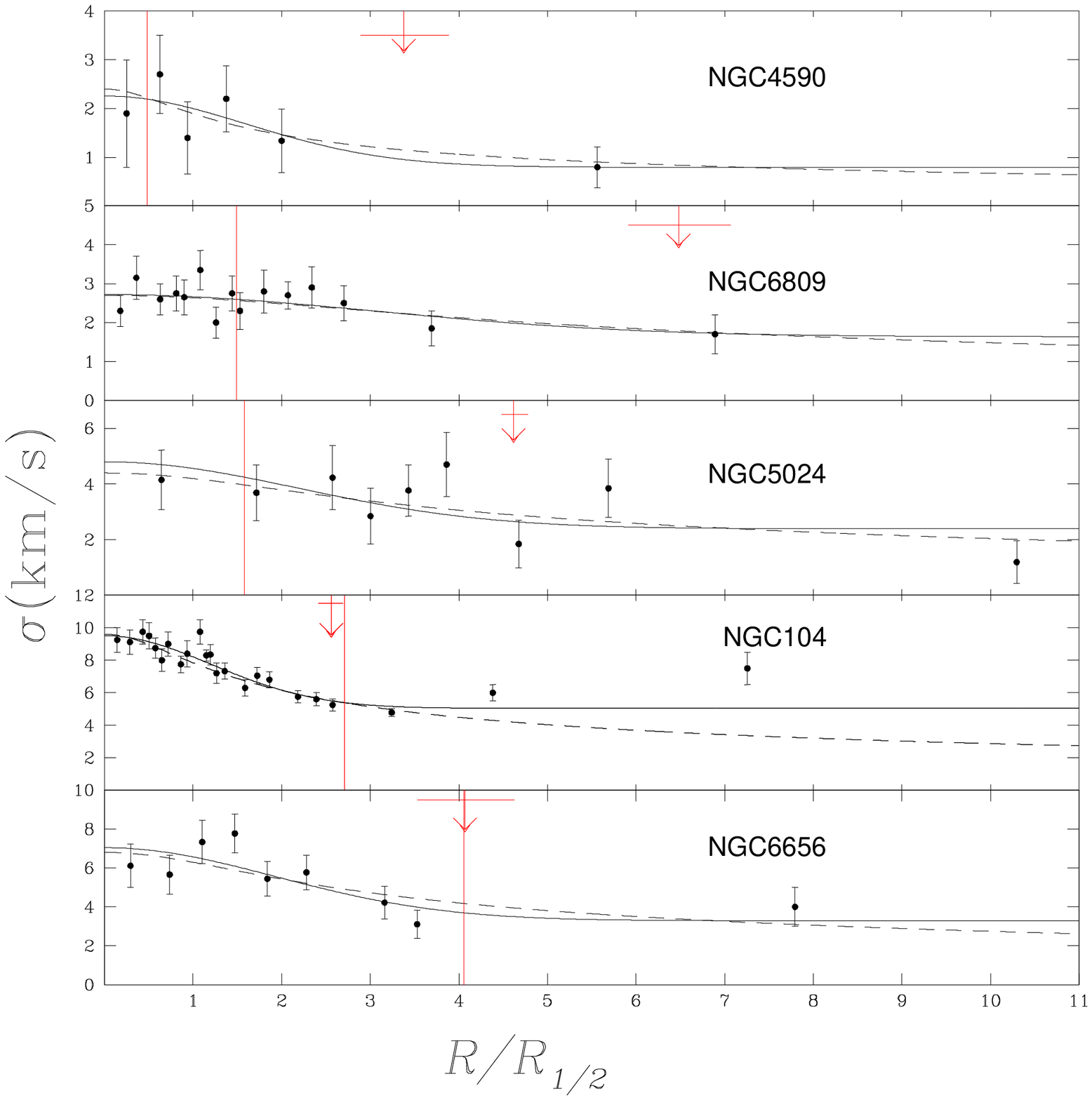}
\caption{The figure shows the observed projected velocity dispersion profiles for five GCs in the Lane et al. sample,
points with error bars, as a function of the radial coordinate, normalised to the half-light radius of each. The solid curves 
give the maximum likelihood fits to the asymptotically flat $\sigma(R)$ model of eq.(1), seen to be accurate descriptions of the data. 
The dashed lines give the best fit Plummer models from the Lane et al. papers, also fair representations of the data. 
The vertical lines indicate the $a=a_{0}$ threshold, and the arrows the point where the profiles flatten, {\it a priori} 
independent features, in most cases seen to occur at approximately the same { region}; see text for details.}
\end{figure*}

\begin{figure*}
\hskip -10pt \includegraphics[width=18cm,height=19cm]{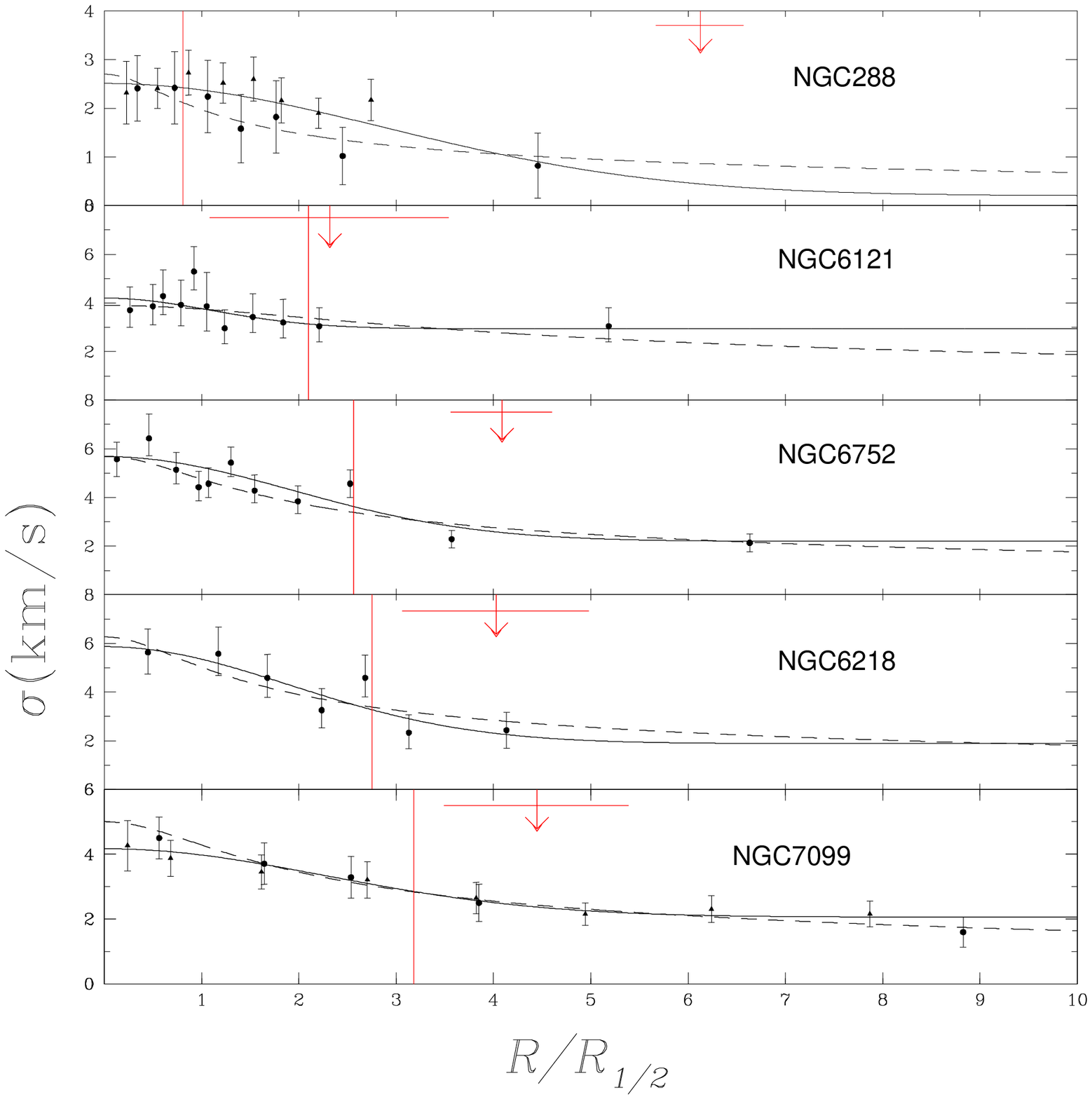}
\caption{The figure shows the observed projected velocity dispersion profiles for the remaining five GCs in the Lane et al. sample,
points with error bars, as a function of the radial coordinate, normalised to the half-light radius of each. The solid curves 
give the maximum likelihood fits to the asymptotically flat $\sigma(R)$ model of eq.(1), seen to be accurate descriptions of the data. 
The dashed lines give the best fit Plummer models from the Lane et al. papers, also fair representations of the data. 
NGC 288 and NGC 7099 are common to both samples, dots and triangles show the { Lane et al. and  Scarpa et al.} data, respectively.
The vertical lines indicate the $a=a_{0}$ threshold, and the arrows the point where the profiles flatten, {\it a priori} 
independent features, in most cases seen to occur at approximately the same { region}; see text for details.}
\end{figure*}

\begin{equation}
\sigma(R)= \sigma_{1} e^{-(R/R_{\sigma})^{2}} + \sigma_{\infty}
\end{equation}

In the above equation $\sigma_{\infty}$ is the asymptotic value of $\sigma(R)$ at large radii, $R_{\sigma}$ a scale radius fixing how
fast the asymptotic value is approached, and $\sigma_{1}$ a normalisation constant giving $\sigma(R=0)=\sigma_{1}+\sigma_{\infty}$.

We now take the observed data points $\sigma_{obs}(R_{i})$ along with the errors associated to each data point, to determine
objectively through a maximum likelihood method the best fit values for each of the three parameters in equation (1), for each 
of the 16 observed globular clusters. Assuming the errors to have a Gaussian distribution, the likelihood function will be:

\begin{equation}
{\cal L} \left( \sigma_{\infty}, \sigma_{1}, R_{\sigma}; \sigma_{obs}(R_{i}) \right) = \prod_{i=1}^{n} 
\frac{ exp[-(\sigma_{obs}(R_{i})-\sigma(R_{i}))^{2}/2 \Delta_{i}^{2}]}{\Delta_{i}}
\end{equation}

\noindent where $\Delta_{i}$ is the error on the i-th data point, and $\sigma(R)$ is a particular model resulting from a
given choice of the three model parameters. Thus, for any choice of the three model parameters, the likelihood function
can be calculated for a given data set $\sigma_{obs}(R_{i})$ with its errors. For each observed globular cluster, we calculate the likelihood
function over a $100^{3}$ grid in parameter space, and then select the point where this function is maximised, to identify
the optimal set of parameters for each observed velocity dispersion profile, $(X_{1,0}, X_{2,0}, X_{3,0})$. As it is customary, 
we work with the logarithm of the likelihood function. The confidence intervals for each of the three parameters are then 
obtained by looking through the full likelihood matrix to identify the largest and smallest values for a particular parameter 
which satisfy the condition $ln{\cal L}(X_{lim}, X_{2}, X_{3}) - ln{\cal L}(X_{1,0}, X_{2,0}, X_{3,0}) =0.5$, i.e., the full projection
of the error ellipsoid is considered, without imposing any marginalisation. This last point allows to properly account for
any correlations between the three fitted parameters when calculating any quantity derived from combinations of them, as will
be constructed in what follows.

Taking $\sigma_{obs}(R_{i})$ data from Drukier et al. (1998), Scarpa et al. (2004), (2007a), (2007b), (2010) and (2011),
Lane et al. (2009), (2010a), (2010b) and (2011) and half-light radii, $R_{1/2}$, from integrating the surface density brightness profiles of Trager et al. 
(1995), we perform a maximum likelihood fit as described above for all the sixteen globular clusters studied. 

Figure (1) shows the observed projected velocity dispersion profiles for the 6 
globular clusters from the Scarpa et al. group, which are not also part of the Lane et al. total sample, points
with error bars. The radial coordinate has been normalised to the $R_{1/2}$ radius of each of the clusters.
The continuous curves show the maximum likelihood fits for each cluster, which are clearly good representations
of the data. We can now give $R_{f}=1.5R_{\sigma}$ as an adequate empirical estimate of the radius beyond which the 
dispersion velocity profile becomes essentially flat. In terms of equation (1), which can be seen to be highly consistent 
with the observed velocity dispersion profiles, $R_{f}$ is the radius such that $\sigma(R_{f})=0.1\sigma_{1}+\sigma_{\infty}$,
a good representation of the transition to the flat behaviour, as can be checked from figure (1), where the arrows
give $R_{f}$, with the horizontal lines on the arrows showing the $1\sigma$ confidence intervals on these fitted parameters.
An empirical definition of the radius where the typical acceleration felt by stars drops below $a_{0}$ can now
be given as $R_{a}$, where:

\begin{equation}
\frac{3 \sigma(R_{a})^{2}}{R_{a}}=a_{0}.
\end{equation}

Using the above definition, we can now identify $R_{a}$ for each of the globular clusters studied. The vertical lines in Figure(1)
show $R_{a}$ for each cluster, also normalised to the half-light radius of each. In the figure, clusters have been ordered by their
$R_{a}/R_{1/2}$ values, with the smallest appearing at the top, and $R_{a}/R_{1/2}$ growing towards the bottom of the figure.

The Lane et al. sample comprises 10 clusters, two of which are also part of the Scarpa et al. sample. Figure 2 is analogous
to Figure 1, and shows velocity dispersion profiles for 5 clusters from the Lane et al. sample not having any overlap with
the Scarpa et al. group. Here we have added also the best fit Plummer models to the data, with parameters taken from the Lane et al. papers,
and shown by the dashed curves. It is obvious that both functional forms provide good representations to the data, which qualitatively,
display an asymptotically flat region at large radii. At large radii, the line of sight velocity dispersion profiles for the Plummer models 
fall to zero, but only very slowly, as $R^{-1/2}$. This allows good fits to data which qualitatively tend to constant values. The good fits
allowed by the Plummer models are clearly not sufficient to dismiss a modified gravity interpretation, as the asymptotically flat projected dispersion 
velocity fits of the type used for full dynamical modelling under modified gravity (Hernandez \& Jimenez 2012) are equally consistent with 
the data. 
 
Figure (3) completes the fits to the Lane et al. sample, where NGC 6121, NGC 6218 and NGC 6752 are analogous to the ones shown in Figure 2. Again, the 
dashed and solid curves are essentially equivalent. The remaining two clusters in this figure give the two examples which have been
studied by both groups of observers, NGC 7099 and NGC288, where the triangles and dots with error bars give the Scarpa et al. and Lane et al.
data, respectively. In these last two cases, the dashed lines give the best fit Plummer models from the Lane et al. papers, and the solid lines,
the best fit models from eq. (1), considering joint data samples from both groups. An eq.(1) fit limited to the Lane et al. data for these last two
clusters was also performed, for the comparison shown in the following figure. For the last two clusters, we see that the two independent
data samples are consistent with a fixed underlying distribution, and also, that the fits to the added samples from eq.(1), represent the data
as well as the Plummer models.

\begin{figure}
\vskip -37pt
\hskip -10pt \includegraphics[height=7.0cm,width=8.5cm]{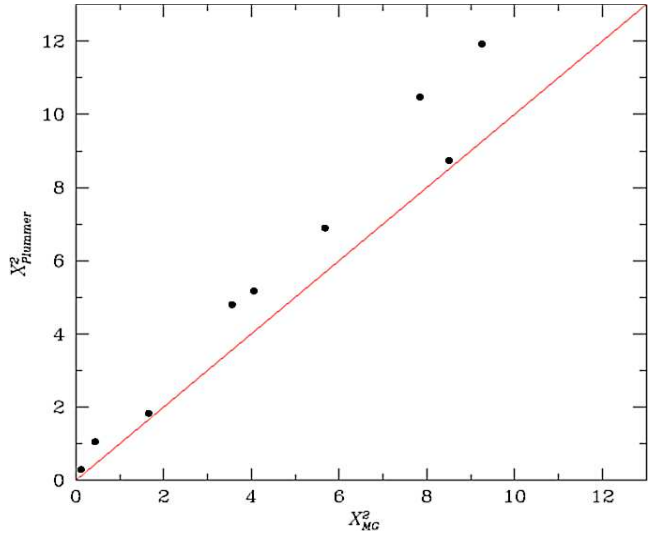}
\caption{The figure shows a comparison of the $\chi^{2}$ values for the best fit asymptotically flat $\sigma(R)$ model of eq.(1),
and the optimum Plummer model fits from the Lane et al. papers to the same data samples. Although both fits are comparable as 
representations of the observed projected velocity dispersion profiles, the asymptotically flat model suggested by modified gravity 
schemes provides, in all cases, a slightly better description of the data.}
\end{figure}

It is interesting at this point to notice a first correlation, the smaller the value of $R_{a}/R_{1/2}$, the larger the
fraction of the cluster which lies in the $a<a_{0}$ regime, and interestingly, the flatter the velocity dispersion
profile appears. At the top of the figures we see clusters where stars experience accelerations below $a_{0}$ almost at all radii, 
and it so happens, that it is only in these systems that the velocity dispersion profile appears almost flat throughout. 
Towards the bottom, we see systems where only at the outskirts accelerations fall under $a_{0}$. Over most of their extents, 
these clusters lie in the Newtonian $a>a_{0}$ regime, and indeed, it is exclusively these, that show a clear Keplerian
decline in the projected velocity dispersion profiles over most of their extents. Also, notice that { on average}, $R_{f}$ and $R_{a}$ 
approximately coincide, as already previously noticed by Scarpa et al. (2007a), the flattening in the velocity dispersion 
profiles seems to appear on crossing the $a_{0}$ threshold.

We end this section with Figure (4) which compares the $\chi^{2}$ values for the Plummer fits to the Lane et al. data, to the
$\chi^{2}$ values for the eq.(1) fits to the same data. As it was already obvious from figures (2)-(3), both functional forms
provide fits of very similar quality, although a rigorous statistical assessment actually shows the fits to profiles which
are asymptotically flat at large radii to better represent the data than the Plummer models, which slowly tend to zero. 
{ NGC 104, which results in the poorest fits under both functional forms tested,  falls off the range shown in Figure (4). 
For the  asymptotically flat profile suggested by MONDian gravity schemes, this cluster yields a $\chi^{2}$ value of 28.55, 
while for the Newtonian Plummer profile of Lane et al., a $\chi^2$ of 51.14 results. This cluster is in fact the one for which 
the difference in $\chi^{2}$ values is greatest, in the sense of further supporting the conclusions presented, but was ommited
from the figure to allow greater detail in the region where the majority of the clusters lie.}

\section{Testing the Newtonian explanation}

In order to test the validity of the explanation for the outer flattening of the observed velocity dispersion profiles under 
Newtonian gravity, that these indicate dynamical heating due to the tides of the Milky Way system (bulge plus disk plus dark halo),
we need accurate estimates of the Newtonian tidal radii for the clusters studied.  One of us in Allen et al. (2006) and 
Allen et al. (2008) performed detailed orbital studies for 54 globular clusters for which absolute proper motions and line of sight
velocities exist. In that study, both a full 3D axisymmetric Newtonian mass model for the Milky Way and a model incorporating a galactic bar 
were used to compute precise orbits for a large sample of globular clusters, which fortunately includes
the 16 of our current study. The Galactic mass models used in those papers are fully consistent with all kinematic and structural
restrictions available. Having a full mass model, together with orbits for each globular cluster, allows the calculation of the
Newtonian tidal radius, not under any ``effective mass'' approximation, but directly through the calculation of the derivative 
of the total Galactic gravitational force, including also the evaluation of gradients in the acceleration across the extent of the 
clusters, at each point along the orbit of each studied cluster. 

The Newtonian tidal radii we take for our clusters, $R_{T}$, are actually the values which result in the largest dynamical heating 
effect upon the clusters studied, those at perigalacticon. As the distance of closest approach to the centre of the Galaxy might 
vary from passage to passage, as indeed it often does, detailed orbital integration is used to take $R_{T}$ as an average
for perigalactic passages over the last 1 Gyr.

\begin{figure}
\vskip -37pt
\hskip -10pt \includegraphics[height=7.0cm,width=8.5cm]{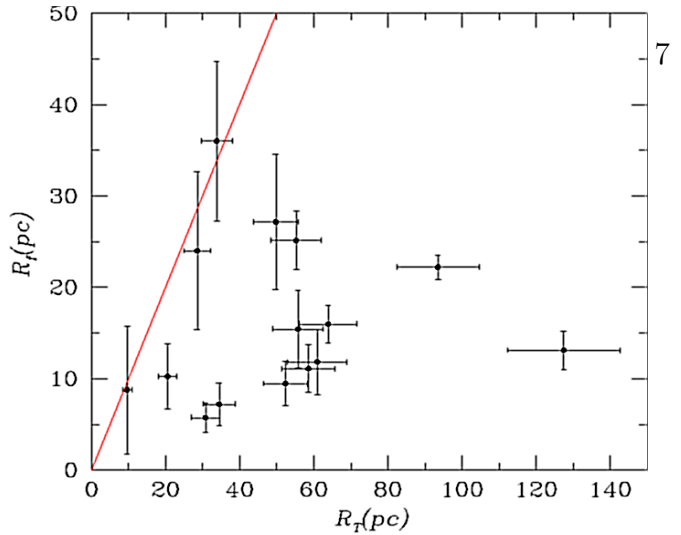}
\caption{The figure shows the relation between the point where the velocity dispersion flattens, $R_{f}$, and the Newtonian tidal radius,
$R_{T}$, for each cluster. Even considering the large errors involved on both quantities, { on average} points fall far to the right of the
identity line shown, making the Newtonian explanation for the flattened velocity dispersion profiles, rather suspect.}
\end{figure}

We update the tidal radii published in Allen et al. (2006) and Allen et al. (2008),
by considering revised total masses from the integration of the observed $V$ band surface brightness profiles for our clusters
(Trager et al. 1995), and using the $V$ band stellar $M/L$ values given in McLaughlin \& van der Marel (2005) and accompanying electronic tables. 
For each individual GC, detailed single stellar population models tuned to the inferred ages and metallicities of each of the 
clusters we model were constructed in that study, using various standard population synthesis codes, and for a variety of assumed 
IMFs. In this way present stellar $M/L$ values in the $V$ band were derived, which we use here. As we do not in any way use the dynamical
mass estimates of McLaughlin \& van der Marel (2005), the total masses we use are independent
of any dynamical modelling or assumption regarding the law of gravity, as they are derived through completely independent surface brightness profile
measurements and stellar population modelling. The confidence intervals in our tidal 
radii include the full range of stellar $M/L$ values given by McLaughlin \& van der Marel (2005), through considering a range
of ages, metallicities and initial mass functions consistent with the observed HR diagrams for each cluster. These uncertainties
dominate the error budget on $R_{T}$, as those introduced by the observational uncertainties in the orbital determinations are much smaller.
This last can be seen from the range in $R_{T}$ values given in Allen et al. (2006) which are extreme in being derived from taking all four
orbital parameters at their $1\sigma$ extremes, something with a probability of $(0.318)^{4}$=1\%, and are hence about $2.58\sigma$
ranges. Although sub-dominant, the corresponding $1\sigma$ errors on $R_{T}$ have also been added. In what follows, we shall make use
of total stellar masses derived as explained above, including as confidence intervals the full uncertainties in these results  
associated with the various IMFs assumed by the McLaughlin \& van der Marel (2005), and not the much narrower confidence intervals resulting 
from taking a fixed IMF.

In Figure (5) we show values of $R_{f}$ for our clusters, plotted against their corresponding $R_{T}$ values, both in units of
pc. The error bars in $R_{f}$ come from the full likelihood analysis described in the fitting process of equation (1) to 
$\sigma_{obs}(R)$, which guarantees that confidence intervals in both of the quantities plotted are robust $1\sigma$ ranges.
The solid line shows a $R_{f}=R_{T}$ relation. It is obvious from the figure that the onset of the flat velocity dispersion regime 
occurs at radii substantially smaller than the tidal radii, for all of the globular clusters in our sample. Even under the most 
extreme accounting of the resulting errors, only three of the clusters studied are consistent with $R_{T} \approx R_{f}$ at $1\sigma$. Actually, 
the average values are closer to  $R_{T}=4 R_{f}$, with values higher than 8 appearing. One of the clusters, NGC 5024 does not appear, as 
it has values of $R_{T}=184.12$, $R_{f}=36$, which puts it out of the plotted range, but consistent with the description given above.  
Given the $R^{3}$ scaling of Newtonian tidal phenomena, even a small factor of less than 2 inwards of the tidal radii, tides can be 
safely ignored, e.g. in Roche lobe overflow dynamics, the stellar interior is largely unaffected by the tidal fields, until almost 
reaching the tidal radius. It therefore appears highly unlikely under a Newtonian scheme, that Galactic tides could be responsible 
for any appreciable dynamical heating of the velocity dispersion of the studied clusters.

We note that Lane et al. (2010) and Lane et al. (2012) find that Newtonian tidal heating can explain the observed velocity dispersion 
profile of their GC sample. However, it is important to note that in Lane et al. (2010)  and Lane et al. (2012), total masses were calculated 
directly from the observed velocity dispersion observations, under the assumption that Newtonian dynamics hold. If that assumption is to be tested, 
the importance of deriving total masses through an independent method, not based on stellar dynamics, is evident. Our results do not imply that the 
Newtonian explanation might not apply to other GCs, e.g. the theoretically constructed ones of K\"{u}pper et al. (2010) and K\"{u}pper et al. (2012),
which show the Newtonian explanation to hold {\it in principle}, although no real GCs were included in those studies. A potential caveat of all the above
studies is the use of strictly axisymmetric potentials for the Milky Way, given that in Allen et al. (2006) and Allen et al. (2008) it is shown
that the orbital dynamics of Galactic globular clusters with orbits probing the regions where the Galactic bar is present, as many in our sample do,
can be strongly affected by its presence.

Notice also, that most of the clusters in our sample are problematic for 
a Newtonian gravity scheme, even without the recent observations of an outer flat velocity dispersion profile. As remarked
already in Allen et al. (2006), the clusters in our sample have Newtonian tidal radii larger than the observed truncation radii
of their light distribution, the sole exceptions being Omega Cen (NGC 5139) and M92 (NGC 6341), two rather anomalous clusters.
Whereas a full dynamical modelling under an extended gravity force law of these clusters, Hernandez \& Jim\'enez (2012), naturally 
yielded an outer truncation for the light profile, under a Newtonian hypothesis, the observed truncation in the light profile of 
the clusters in our sample cannot be explained as arising from interaction with the tidal field of the Milky Way.

Furthermore, notice that we have taken $R_{T}$ at perigalacticon, where tides are at their most severe over the clusters orbit, any 
other orbital occupation averaging would result in substantially larger $R_{T}$ values. Notice also that as shown by Allen et al. (2006) 
and Allen et al. (2008), the inclusion of a realistic massive Galactic bar potential, in the case of the clusters in our sample, results 
generally in negligible changes in the resulting $R_{T}$ values, or in some cases, a slight increase in these values. Hence, even taking 
the fullest non-axisymmetric Galactic mass model under Newtonian gravity, with precise orbits derived from 3D velocity measurements for 
the clusters studied, together with total mass determinations tuned to the individual stellar populations of them, yields tidal radii as 
shown in figure (2).

Regarding a comparison to the expectations under Newtonian gravity, an interesting dynamical effect appears when a stellar halo object
is near its apocentre. As shown in e.g. Niederste-Ostholt et al. (2012), near apocentre tidal tails are compressed into what might 
look like a high dispersion velocity halo. We have checked the position of the clusters studied along their orbits, and found 
that only in the case of NGC 6121 is the cluster near apocentre, checked explicitly from the orbits for the clusters in question 
from Allen et al. (2006) and Allen et al. (2008). Notice that this case also follows the MONDian expectations of figure (7), see below.
It is also important to note that the piling up of tidally stripped stars near 
apocentre has not been proven to hold for more chaotic orbits, and probably does so to a much smaller degree than what shown in 
Niederste-Ostholt et al. (2012) for a pure axisymmetric potential. This is relevant, as the orbits of clusters lying within the region 
of influence of the galactic bar, as many of the ones in our sample do, become substantially chaotic, with no well defined periodic 
apocentre distance, as shown in the Allen et al. papers mentioned.

\section{Testing a Modified Gravity explanation}

\begin{figure}
\hskip -10pt \includegraphics[height=7.0cm,width=8.5cm]{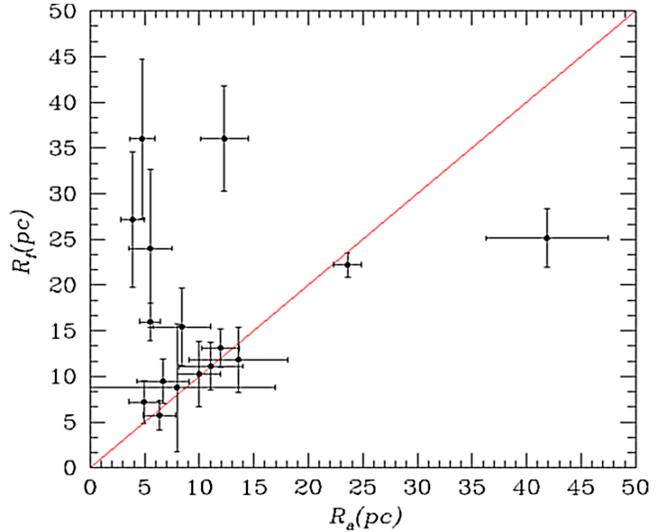}
\caption{The figure gives the relation between the radius where the velocity dispersion flattens, $R_{f}$, and the point
where average stellar accelerations fall below the $a_{0}$ threshold, $R_{a}$. { }}
\end{figure}

We begin this section by testing the correlation between $R_{a}$ and $R_{f}$. As already noticed by Scarpa et al. (2011), the flattening
in the observed velocity dispersion profiles seems to appear at the point where the $a_{0}$ threshold is crossed. Here we use the much more
careful and objective modeling of the observed velocity dispersion curves of the previous section to test this point, shown in Figure (6).

{ We see 9 GCs in the sample falling within $1\sigma$ of the identity line shown, a further three lying within $2\sigma$ of this same line,
and the remaining four appearing as outliers. Thus, the correlation appears stronger than in Figures (1)-(3), where errors on $R_{a}/R_{1/2}$
appear and only a qualitative comparison
is implied, not including the confidence intervals in $R_{a}$.} A quantitative test of the correlation being explored is possible, since the careful modelling 
of the velocity dispersion profiles we performed naturally yields objective confidence intervals for the parameters of the fit.
Of the outliers, NGC288 presents an almost entirely flat velocity dispersion profile, and is hence a case where
the parameter $R_{\sigma}$ is only poorly constrained. In this figure we thus quantify the correlation between the point where the velocity
profile flattens, and the crossing of the $a_{0}$ threshold, as expected under MONDian schemes, which is seen to hold { on average}. One could think 
of adding a point at (0,0) in Figure (6), corresponding to the local dSph galaxies, systems with fully isothermal observed velocity dispersion profiles, 
lying fully within the $a<a_{0}$ condition, e.g. Angus (2008), Hernandez et al. (2010). { A possible caveat is the use of only a coarse definition 
for Ra, which relies only on projected quantities which are integrals along the line of sight. More detailed dynamical structure modelling of the type 
found in Hernandez \& Jim\'enez (2012), requiring fixing on a particular modified gravity model, something which we expressedly avoid in the present study,
might reveal slight differences from the current Figure (6), perhaps with no outliers.}

\begin{figure}
\hskip -10pt \includegraphics[height=7.0cm,width=8.5cm]{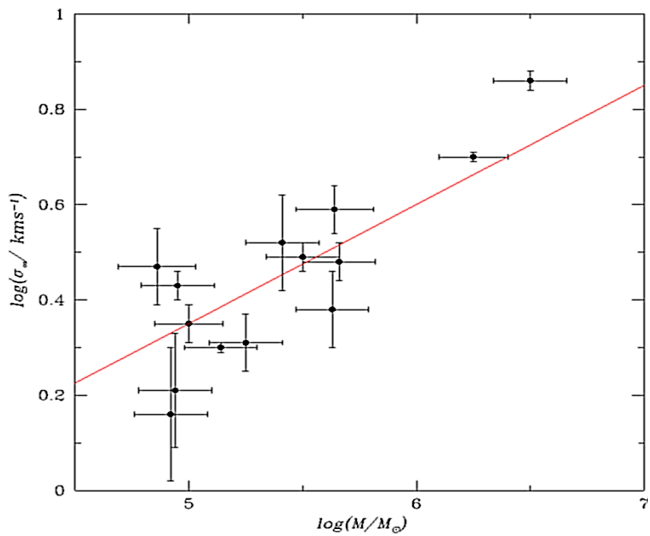}
\caption{Here we give the relation between the observed asymptotic dispersion velocity measurements, and the total mass of each
cluster. The line gives the best fit $\sigma \propto M^{1/4}$ scaling for the data, and is consistent with the galactic scale Tully-Fisher 
relation.}
\end{figure}

As already noticed in Figures (1)-(3), a correlation is evident in that the further out, in units of the cluster half-light
radius, that the $a_{0}$ threshold is reached, the larger the relative drop in the observed velocity dispersion profile. This is
expected under MONDian gravity schemes, since when the Newtonian $a>a_{0}$ region is larger within a particular cluster, 
the larger the ``Keplerian'' fall before the $a<a_{0}$ modified regime is reached.

\begin{table*}
\begin{flushleft}
  \caption{Parameters for the globular clusters treated.}
  \begin{tabular}{@{}lllllllllll@{}}
  \hline
 \hline
   GC &   $\sigma_{1} (km/s) $ & $\sigma_{\infty} (km/s)$  & $R_{\sigma} (pc)$ & $R_{a} (pc)$ & $R_{1/2} (pc)$  &  $R_{T} (pc)$  & $log_{10}(M/M_{\odot})$ & \,\, $b$ & $\Delta V_{r} (km/s)$\\
 \hline
 & & & & & \\

 NGC 104$^{2}$       & $4.5 \pm 0.4$   & $5.0 \pm 0.1$ & $14.8 \pm 0.9$  & $23.6 \pm 1.3$  & $8.7 \pm 1$ & $93.5 \pm 11.1$ &  $6.3 \pm 0.2$  & -45.20  &  -18.7\\
 & & & & & \\
 NGC 288$^{1,2}$     & $2.3  \pm 1.4$  & $0.2 \pm 0.8$ & $24.0 \pm 5.8$  & $4.8 \pm 1.2$   & $5.9 \pm 1$ & $33.8 \pm 4.2$ &  $4.8 \pm 0.2$   & -89.38  &  -46.6\\
 & & & & & \\
 NGC 1851$^{1}$      & $4.2 \pm 3.0$   & $3.9 \pm 0.4$ & $7.9 \pm 2.4$   & $13.6 \pm 4.5$  & $2.4 \pm 1$ & $60.9 \pm 10.7$ &  $5.6 \pm 0.2$  & -35.03  &  320.5\\
 & & & & & \\
 NGC 1904$^{1}$      & $2.4 \pm 0.3$   & $2.0 \pm 0.1$ & $10.3 \pm 2.9$  & $8.4 \pm 2.6$   & $2.7 \pm 1$ & $55.8 \pm 9.4$ &  $4.9 \pm 0.2$   & -29.35  &  206.0\\
 & & & & & \\
 NGC 4590$^{2}$      & $1.5 \pm 0.6$   & $0.8 \pm 0.4$ & $18.1 \pm 5.0$  & $3.9 \pm 1.1$   & $8.0 \pm 1$ & $49.8 \pm 6.0$ &  $5.7 \pm 0.2$   & 35.8    &  -94.3\\
 & & & & & \\
 NGC 5024$^{2}$      & $2.4 \pm 1.5$   & $2.4 \pm 0.5$ & $24.0 \pm 3.8$  & $12.3 \pm 2.2$  & $7.8 \pm 1$ & $184.1 \pm 22.2$ &  $5.6 \pm 0.2$ & 79.3    &  -62.9\\
 & & & & & \\
 NGC 5139$^{1}$      & $8.0 \pm 1.0$   & $7.2 \pm 0.4$ & $16.8 \pm 2.1$  & $41.9 \pm 5.6$  & $7.2 \pm 1$ & $55.3 \pm 8.9$ &  $6.4 \pm 0.2$   & 14.97   &  232.3\\
 & & & & & \\
 NGC 6121$^{2}$      & $1.3 \pm 1.0$   & $2.9 \pm 0.6$ & $5.9 \pm 4.6$   & $8.0 \pm 8.0$   & $3.8 \pm 1$ & $9.6 \pm 1.3$ &  $4.9 \pm 0.2$    & 15.38   &  70.7\\
 & & & & & \\
 NGC 6171$^{1}$      & $1.4 \pm 0.6$   & $2.7 \pm 0.2$ & $3.8  \pm 1.1$  & $6.4 \pm 1.5$   & $3.2 \pm 1$ & $30.8 \pm 5.2$ &  $4.9 \pm 0.2$   & 23.01   &  -33.6\\
 & & & & & \\
 NGC 6218$^{2}$      & $3.3 \pm 0.9$   & $1.4 \pm 0.5$ & $4.7 \pm 1.2$   & $4.9 \pm 1.4$   & $1.8 \pm 1$ & $34.5 \pm 4.4$ &  $4.9 \pm 0.2$   & -42.20  &  25.71\\
 & & & & & \\
 NGC 6341$^{1}$      & $3.8 \pm 0.7$   & $3.1 \pm 0.2$ & $6.8 \pm 2.4$   & $10.0 \pm 2.0$  & $3.2 \pm 1$ & $20.6 \pm 3.4$ &  $5.3 \pm 0.2$   & 34.86   & -120.3\\
 & & & & & \\
 NGC 6656$^{2}$      & $3.8 \pm 1.3$   & $3.3 \pm 0.8$ & $7.4 \pm 1.7$   & $11.1 \pm 2.9$  & $2.7 \pm 1$ & $58.6 \pm 7.0$ &  $5.4 \pm 0.2$   & -8.15   &  -146.3\\
 & & & & & \\
 NGC 6752$^{2}$      & $3.5 \pm 0.8$   & $2.0 \pm 0.3$ & $10.7 \pm 1.4$  & $5.5 \pm 0.9$  & $3.7 \pm 1$  & $63.9 \pm 7.8$ &  $4.9 \pm 0.2$   & -23.87  &  174.7\\
 & & & & & \\
 NGC 6809$^{2}$      & $1.1 \pm 0.8$   & $1.6 \pm 0.5$ & $16.0 \pm 5.8$  & $5.5 \pm 2.0$  & $2.7 \pm 1$  & $28.6 \pm 3.6$ &   $4.9 \pm 0.2$   & -29.35  &  206.0\\
 & & & & & \\
 NGC 7078$^{1}$      & $5.1 \pm 1.6$   & $3.0 \pm 0.3$ & $8.7 \pm 1.4$   & $12.0\pm 1.7$  & $3.2 \pm 1$  & $127.5 \pm 21.0$ & $5.5 \pm 0.2$   & -27.31  & -107.0\\
  & & & & & \\
 NGC 7099$^{1,2}$    & $2.1 \pm 0.4$   & $2.0 \pm 0.2$ & $6.8  \pm 1.1$  & $6.7 \pm 2.4$  & $2.1 \pm 1$  & $53.4 \pm 8.3$ &  $4.8 \pm 0.2$    & -46.80  & -185.0\\
 \hline
\end{tabular} 

The first three entries give the parameters of the fits to the observed projected velocity dispersion profiles and their confidence intervals,
to data from the Scarpa et al. group $^{1}$ and the Lane et al. group $^{2}$. The fourth column gives an empirical estimate of the point where 
the average stellar acceleration drops below $a_{0}$. Columns 5-7 give the half-light radius calculated from the surface density light profiles, 
the Newtonian tidal radius from the Galactic mass model and orbital calculations derived from observed proper motions by Allen et al. (2006), 
and the total masses derived from the $M/L$ values inferred through stellar population modelling by McLaughlin \& van der Marel (2005) for each cluster, 
with corresponding confidence intervals. The last two columns give the galactic latitude of the clusters, and the radial velocity difference 
with respect to the Sun. 
\end{flushleft}
\end{table*}

We end this section with Figure (7), which shows the relation between the measured asymptotic velocity dispersion, $\sigma_{\infty}$,
and the total mass of the clusters in question. The mass was calculated as described in section (3), and therefore represents 
the best current estimate of the stellar mass for each of the clusters in the sample, including its corresponding confidence intervals.
As with all the other correlations and data presented in figures (1)-(6), there is no dynamical modelling or modified gravity
assumptions going into the data presented in Figure (7), merely observable quantities. We see, as already pointed out
in Hernandez \& Jimen\'ez (2012), that the GCs observed {} comply with a scaling of $\sigma \propto M^{1/4}$, the Tully-Fisher
law of galactic systems ``embedded within massive dark haloes''. 

From this last figure two clusters have been excluded, NGC 288 and NGC 4590
which have very poorly determined $\sigma_{\infty}$ values, with uncertainties so large that these clusters provide little information in terms of Figure (7).
{ Regarding NGC 288, both the Scarpa et al. and the Lane et al. data are noisier than for the others. Also, taking only one set of data yields 
significantly distinct answers, although still barely within their respective errors, a low asymptotic velocity for the Lane et al. sample, a high 
value from the Scarpa et al. data. This does not happen with the other common GC, NGC 7099, where both data samples are in complete agreement. 
Although Sollima et al. (2012) recently reported a velocity dispersion profile for NGC 288, this is equally noisy, and does not help to clear the case, 
their data are actually consistent, to their respective errors, with both the Lane et al. or the Scarpa et al. data. Given the current level of observational 
uncertainties we prefer to exclude NGC 288 from any consideration regarding its asymptotic velocity value, until a clearer picture emerges from the 
observational point of view.}

The straight line shows the best fit $\sigma \propto M^{1/4}$ scaling, and actually falls only a factor of 1.3 below the modified gravity 
prediction for systems lying fully within the low acceleration regime (e.g. Hernandez \& Jimen\'ez 2012), for the same value of
$a_{0}=1.2 \times 10^{-10} m/s$ used here, as calibrated through the rotation curves of galactic systems. This small offset is not surprising, 
since the GCs treated here are not fully within the $a<a_{0}$ condition, most have an inner Newtonian region encompassing a substantial 
fraction of their masses. { Notice also that from an statistical point of view, consistency of a set of 
data points with a model does not require for all data points to lie within $1\sigma$ of the proposed model. Probabilistically, one actually 
expects about $1/3$ of the points to lie between $1\sigma$ and $2\sigma$ of the model, with $1/100$ expected between $2\sigma$ and $3\sigma$, even for 
data actually extracted from a given model. Given the size of our sample, finding 5 GCs without $1\sigma$, but within $2\sigma$ of the proposed 
model is then well within expected random noise, inasmuch as we have taken care to ensure that the error bars given are real $1\sigma$ confidence intervals.
Further, as many of the GCs in our sample have not reached acceleration values significantly below $a_{0}$ at their last measured 
point, while other have, a certain intrinsic scatter would be expected in Figure (7) from a MONDian gravity perspective.

From a Newtonian point of view, if Galactic tides were responsible for the observed outer flattening of the velocity dispersion profiles
studied, given the narrow range of half light radii these present, and given the inverse scaling of Newtonian tides with the density
of the satelite, a slight downward trend for decreasing asymptotic velocity dispersion with increasing mass would be expected in Figure (7).
This would of course be blurred significantly by the range of perigalacticon distances inferred for the GCs in our sample. It is clear
from the figure that a blurred decreasing trend is not what the data show; rather, consistency with the $\sigma \propto M^{1/4}$ 
of MONDian gravity, including the normalisation, is evident.} A preliminary version of this last figure appeared already in Hernandez \& Jim\'enez (2012); 
we reproduce here an updated version using now the extended sample of clusters treated, and $\sigma_{\infty}$ values and their confidence intervals as 
derived through the careful velocity dispersion fitting procedure introduced.

Given that the inclusion of even a few high-velocity contaminating stars can bias the velocity dispersion measurements significantly,
e.g. Giersz \& Heggie (2011), it is important to assess the robustness of the velocity dispersion profiles we use to this
possibility. The stars from the Lane et al. group  were selected through the requirement of four stellar parameters, 
ensuring membership through requiring simultaneously a $\Delta V_{r}$, and also $Ca$, $g$ and $[m/H]$ membership criteria. These makes
it unlikely that contamination issues might have degraded the velocity dispersion profiles reported by
the Lane et al. group. Regarding the Scarpa et al. results, only a $\Delta V_{r}$ membership criteria was used. However, as can be seen
from the table, only one of their clusters, NGC 6171, has $|b|<45$ and $\Delta V_{r}<100 km/s$, showing that with this only possible 
exception, contamination of field stars is unlikely to affect the derived velocity dispersion determinations we use.
It is also reassuring of the reliability of the Scarpa et al. results, that of the two clusters which have also been studied by the Lane et al. 
group, NGC 7099 has reported velocity dispersion profiles which are fully consistent { when comparing the data samples from the two groups of observers.  
The case of NGC 288 has already been discussed, although both data samples are still consistent to within their respective errors, a definitive trend appears 
for large and small asymptotic velocity dispersion values for the Scarps et al. and Lane et al. groups, respectively}

To summarise, we have tested the Newtonian explanation of Galactic tides as responsible for the observed $\sigma(R)$ phenomenology,
and found it to be in tension with the observations, given the tidal radii (at perigalacticon) which the GCs in our sample present, are generally 
larger than the points where  $\sigma(R)$ flattens, on average, by factors of 4, with values higher than 8 also appearing. An explanation under a 
MONDian gravity scheme appears probable,
given the { } correlations we found for the clusters in our sample, all in the expected sense, and shown in Figures (1)-(7). Table (1) gives the 
parameters of velocity dispersion fits and their confidence intervals. The errors in $\sigma_{\infty}$ are uncorrelated with those in the other two 
parameters, which as it is easy to see from the model, are perfectly anti-correlated amongst themselves. The masses come from integrating the observed 
surface density light profiles, and using the $M/L$ values, and their uncertainties, calculated using detailed stellar population modelling on a cluster 
by cluster basis by McLaughlin \& van der Marel (2005).

\section{Conclusions}\label{ccl}

From a purely empirical perspective, we test the Newtonian explanation of Galactic tides as responsible for the observed
flattening of the velocity dispersion profiles in the GCs studied. These clusters can be shown to have Newtonian tidal radii
at closest galactic passage larger than the points where $\sigma(R)$ flattens, by large factors of 4 on average, making the
explanation under the Newtonian hypothesis suspect.

Through a careful modelling of the observed velocity dispersion profiles, we corroborate { an average} correlation between 
the appearance of a flat region in  $\sigma(R)$ and the crossing of the $a_{0}$ threshold, as expected under modified gravity schemes.

By including results from careful stellar population modelling of the globular clusters studied to derive total mass estimates,
we show that the asymptotic values of the measured velocity dispersion profiles, $\sigma_{\infty}$, and total 
masses for these systems, $M$, are consistent with the generic modified gravity prediction for a scaling 
$\sigma_{\infty}^{4} \propto M$.

{
Although individual velocity dispersion profiles can be adequately fitted with either Newtonian Plummer models, or MONDian asymptotically 
flat ones to equivalent accuracy, the large Newtonian tidal radii sometimes found and the "Tully-Fisher" mass-velocity scaling observed,
show that the phenomenology of the velocity dispersion profiles of the globular clusters studied is consistent with a qualitative 
change in gravity in the low acceleration regime, as predicted by MONDian gravity theories.
}

\section*{acknowledgements}

The authors thank an anonymous referee for a thorough reading of a previous version of the paper, leading to a helpful report abundant 
in constructive criticism. Xavier Hernandez acknowledges financial assistance from UNAM DGAPA grant IN103011. Alejandra Jim\'enez 
acknowledges financial support from a CONACYT scholarship.

\end{document}